# A Survey of Data Marketplaces and Their Business Models

Santiago Andrés Azcoitia
santiago.azcoitia@imdea.org
IMDEA Networks Institute
Universidad Carlos III Madrid
Leganés, Madrid, Spain

Nikolaos Laoutaris
nikolaos.laoutaris@imdea.org
IMDEA Networks Institute
Leganés, Madrid, Spain

## ABSTRACT

"Data" is becoming an indispensable production factor, just like land, infrastructure, labor or capital. As part of this, a myriad of applications in different sectors require huge amounts of information to feed models and algorithms responsible for critical roles in production chains and business processes. Tasks ranging from automating certain functions to facilitating decision-making in data-driven organizations increasingly benefit from acquiring data inputs from third parties. Responding to this demand, new entities and novel business models have appeared with the aim of matching such data requirements with the right providers and facilitating the exchange of information. In this paper, we present the results and conclusions of a comprehensive survey on the state of the art of entities trading data on the internet, as well as novel data marketplace designs from the research community.

## 1 INTRODUCTION

Paying for information is not a new idea: insiders have been hired and spies have been trained to achieve a competitive advantage while doing business or fighting wars since ancient times. Such primitive information exchanges exclusively involved humans, and would sometimes result in the death of the messenger, yet they were often decisive and undeniably influenced the course of history (e.g., Ephialtes betrayal in the Battle of Thermopilae).

Later, with the advent of telecommunications, information was no longer transmitted by people but by electromagnetic signals, and the exchange of information became instantaneous. Later still, computing, electronics and digital communications gave birth to a new generation of sensors and increasingly automated data collection. As a result, the majority of information now flows from machines to humans.

An even more revolutionary twist will likely drive the future growth of the so-called knowledge economy thanks to the internet of things (IoT), artificial intelligence (AI), and ubiquitous communication systems such as 5G. According to IDC, 30% of data will be generated by sensors in real time by 2025 [53]. In the current context of the major digitalization of the economy, a myriad of applications and data-hungry machine learning (ML) models are - to give a couple of meaningful examples - helping companies and public institutions improve their efficiency, as well as assisting individuals in health issues. This means that machines will join humans as the main data consumers. In some settings, such M2M data exchanges will be required to happen in real time, too.

The global amount of new data created every year will grow 530% from 2018 to 2025 [53]. Not only is an increasing digitalization expected to cause a dramatic ramp up in the volume of data exchanged, but an unparalleled rise in the size of the data economy, too. A McKinsey report predicted that the data-driven decision-making market will reach US$2.5 trillion globally by 2025 [43], whereas a recent study within the scope of the European Data Strategy estimates a size of 827 billion euro for the EU27 [23], close to 80% of the yearly GDP of a country like Spain. Regardless of the fact that "data" is a commodity like oil, capital, an asset, or similar to labor [7], it is undoubtedly becoming a cornerstone of the knowledge economy in the 21st century.

Varying entities have recently had to respond to the exponential increment in the demand for data. Traditionally, data providers have long collected and enriched public information scraped from the internet and their users. Leveraging those valuable and (in time) reputable information silos, they have built successful business models mainly around marketing (Acxiom, Experian, etc), or financial or business intelligence (Bloomberg, Thomson Reuters, etc.). More recently, data marketplaces (DMs) - two-sided platforms intended to match data sellers and buyers and, in some cases, facilitate and manage data exchanges and transactions - have also arrived on the scene.

First-generation *general-purpose* DMs are being complemented by *niche* DM platforms that target specific industries (e.g., Caruso for the connected car, Veracity for energy and transportation), and cover data sourcing for specific innovative purposes, such as feeding AI / ML algorithms (Mechanical Turk, DefinedCrowd), or trading IoT real-time sensor data (IOTA, Terbine).

Not surprisingly, some leading data-management systems (e.g., Snowflake, Cognite) and niche digital solutions (e.g., Carto, Openprise, LiveRamp) are integrating secure data exchanges and, in some cases, enabling an internal data market to buy and sell data within the system. Such *embedded private* marketplaces provide their users with a fit-for-purpose complementary sourcing functionality to quickly find and seamlessly integrate useful data from third parties.

Along with an increasing concern about online privacy, some start-ups have developed innovative solutions to manage and monetize personal data from individuals in the last decade. Such Personal Information Management Systems (PIMS) have been spurred by recent legislative developments, such as the General Data Protection Regulation (GDPR) in the EU or the California Consumer Privacy Act (CCPA) in the US. Leveraging such legislation, PIMS empower individuals to take control of their personal information (PI) made available to internet service providers, and to manage their consent so that their data is only given away to certain entities, or for some specific purposes. Moreover, some PIMS have also implemented marketplace functions for users to sell their consent at a price, hence enabling the monetization of PI.

Due to the nature of data as an asset (freely replicable, non-perishable, serving a wide range of uses, holding an inherently combinatorial and aprioristically unknown value, which also depends on the buyer and the use case [1, 42]), commercial data markets are still immature and do not suffice for realizing the benefits of a widespread information exchange. As a result, most information still remains in silos, and data sharing often requires signing bilateral partnership agreements and *ad hoc* opaque bargaining in practice. In this context, market players usually fight to integrate themselves horizontally into the value chain [56], and secure a niche where they can act as a *de facto* monopoly by leveraging and fiercely protecting their core data, as their main competitive advantage. This structural market fragmentation is ultimately deterring the potential benefits of a healthy FAIR [64] data economy.

We set out to study how different entities are selling data in the market, what kind of relationships are taking place in the value network, how data trading is evolving, and what challenges must be overcome in order to unleash the power of data in the economy.

## 1.1 Our contributions

Despite the increasing importance of data in the economy, the ecosystem around data trading remains largely unknown to the scientific community. In this paper, we list 180 data trading entities (DTEs), and conduct an in-depth study of 97 of them based on public information available on the internet and published by these companies. We answer questions which address issues such as what kind of data they trade, how they collect and manage it, whom they sell such information to, what they provide to them, or how they deliver and price their services, among others. By analyzing this information, we spot and characterize ten different data trading business models, identify relevant market trends, and discuss relevant challenges for DTEs.

Moreover, we also scrape information about more than 210,000 data products and list 2,015 distinct data providers present in twelve public data marketplaces in order to better understand what the most popular categories of data are and who is selling data through which marketplaces.

## 1.2 Related Works

Different directories of DMs are available on the internet [19, 51] - some of them powered by governments [49] - listing commercial DMs and initiatives alike. Though they proved very useful in building the survey base, they do not analyze the different entities in detail, nor do they compare their business models as we do in this paper.

Other survey papers have recently been published regarding data marketplaces [46, 55, 57]. To the best of our knowledge, ours is more up-to-date, broader in scope, and provides an in-depth analysis of three times more entities than previous works. Furthermore - and following our study of nineteen of them - this work is also, to the best of our knowledge, the first to address the business models and challenges of PIMS.

## 1.3 Our Findings

As to the kind of data being traded on the internet, we found, unsurprisingly, that marketing and financial DMs are currently the most popular. Not only do entities focused on marketing and financial data outnumber those trading other categories of data, but, most data products offered in commercial DMs also belong in these two categories. However, most novel proposals are aimed at monetizing real-time IoT data, managing personal information of individuals, and delivering trained AI/ML models - as opposed to providing data for buyers to train them.

In general, we spotted some interesting trends in how

(1) brand-new platforms opt for distributed rather than centralized architectures when storing or processing data.

(2) DMs are progressively using distributed ledger technologies (DLT) for managing, and accounting for transactions.
(3) data transactions are increasingly being paid in cryptocurrency rather than fiat currency which, together with the usage of DLT, is meant to speed up real-time data transactions by avoiding the extra latency caused by financial intermediaries.

Even though every player in the ecosystem faces its own specific challenges, bootstrapping and scaling data-exchange and marketplace platforms represent a common challenge. Therefore, increasing the value they provide and ensuring the trust of every party involved become of paramount importance for new data trading platforms. It became clear to us that DTEs are approaching these challenges by specializing and commodifying data exchanges.

First, we found that new DMs are concentrating their activity in specific industries, or in the types of data they have expertise in. A narrower focus allows them to increase the value they are able to provide to users, be they buyers or sellers (e.g., tailoring their services or involving representative stakeholders of the industry). As a result, the market is evolving from more traditional monolithic general-purpose DMs towards "niche" data trading platforms, which are more likely to grow and become a point of reference within their scope.

With regards to commodification, some market players consider the ability to integrate data from third parties as a functionality of their data-driven products or services. By considering data trading a means rather than the ends of their activity, a DM avoids the problem of bootstrapping new solutions as long as it complements services which are already mature and well established.

Data trading poses other major general challenges such as dealing with ownership and fighting against theft of data, establishing acknowledgeable price references, dealing with market fragmentation, and developing open layered standards to operationalize the secure exchange of data on the internet. Due to the heterogeneity of information 'goods' and their uses, some other relevant challenges apply in specific settings only. This is the case with personal data protection or, in the case of sourcing data for AI/ML models, being able to select the most suitable samples for a particular task, pricing them accordingly, and rewarding data sellers proportionally to their contribution.

The remainder of the paper is structured as follows:

- Section 2 introduces the data value chain and the concept of '*business model*'.
- Section 3 presents the scope of the survey, and characterizes a catalog of business models we found during our study.
- Section 4 provides more detail on the results of our survey and how different entities carry out data exchange and trading on the internet.
- Section 5 presents a state-of-the-art of novel DM proposals from the research community.
- Section 6 points to key relevant challenges and research directions for data trading to become operational.
- Section 7 summarizes the key takeaways from our analysis and presents some trends in this emerging ecosystem.

As a starting point, we provide some background to our study, and introduce some of the terms and definitions we will use throughout the paper, along with our view of the data value chain.

## 2 UNDERSTANDING THE DATA TRADING VALUE CHAIN

In the context of data trading, *actors* in the value chain are legal entities or individuals playing an effective role in producing any data-driven service or data product, be it intermediate or final, that is offered and eventually acquired in the market. We will generally refer to them as data trading entities or DTEs. Our survey aims to understand what the roles of such DTEs are, how they interact with other DTEs, how they do business, and what mechanisms they use to set prices for data. We encapsulate all this information in the concept of a *business model*, a term that has been defined in various ways in the literature [47]. For the purpose of this paper, we will refer to a DTE's *business model* as the description of its value proposition within the chain, the processes or activities it covers, the inputs it requires, and the outputs it provides the market with, as well as the relationship the entity maintains with other actors [16].

Understanding the data value chain is a key first step in order to identify relevant business models. Previous studies have already explained the data value chain in general [17, 29], and specific contexts [40, 41]. From a broad data trading perspective, Fig. 1 shows a diagram of four stacked functional layers that allow sellers and buyers to connect. We will later use this to position and classify actors in the market.

At the bottom, the *infrastructure layer* provides the basic processing, secure storage and communication functions to the upper layers in the stack.

On top of such infrastructure, the *enablement layer* provides generic application programming interfaces (APIs) and functions to DTEs. Some solutions and PDKs in the market do not intend to directly provide services to the end users, but rather to provide a platform with common useful functions that *enable* other DTEs to carry out a controlled data exchange which optionally may involve an economic transaction.

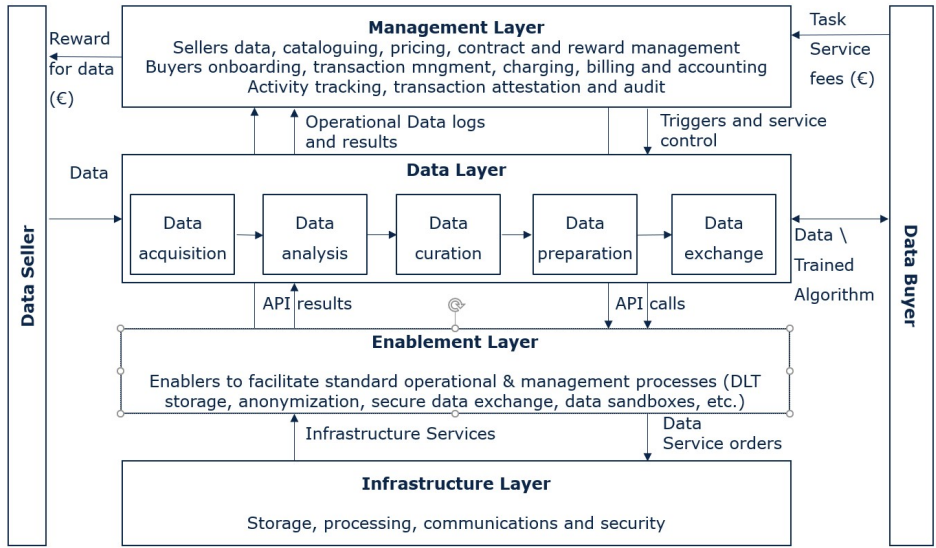

**Figure 1: A layered approach to data trading**

In the next level, a more technical and operational *data layer* deals with data processing itself and responds to the effective delivery of data or data-driven services to end-customers, be they consumers, or another DTE. Reaching from data collection or extraction to its final delivery to the end-consumer, this process usually requires intermediate preprocessing, curation and data enrichment steps. In addition, it may involve third parties whose data is acquired and combined, and therefore other secure data exchanges.

Finally, the top *management layer* deals with data discovery, coordinates transactions, keeps track of contracts and service level agreements, and ensures the accountability and transparency of all the operations and processes in the data layer. In contrast to the operational *data layer* immediately below, it works with metadata and transactional data. Other functions of the management layer include helping data-owners catalogue, structure and price their data offer, governing data transactions (e.g., through contract management, charging, billing and accounting processes), and increasing the overall transparency of data trading. In the case of transactions involving data from multiple sellers, it is also in charge of distributing the resulting payments among them.

Note that our definition allows for cascading transactions, which is oftentimes the case before sufficiently processed data is transformed to a data-driven service to end-consumers. For example, a model that outputs consumer segmentation data at postcode level requires at least the following steps: i) gathering anonymized segmentation data (often from disparate sources), ii) combining such information with geo-located identity data into a single coherent dataset, and iii) aggregating this output into individual postcodes by processing it together with postcode border shapefiles (often obtained from a third party, too) in a geographical information system.

## 3 A TAXONOMY OF DATA TRADING BUSINESS MODELS

### 3.1 The universe of Data Trading Entities

We initially checked more than 180 companies offering data products on the internet. After a brief initial review, we selected 97 of them to analyze in detail. We discarded concept projects, online advertising platforms, and internet service providers not specifically offering data products.

The final set includes companies of different sizes from 22 countries, as Fig. 2 shows. Furthermore, we collected information about when these companies were founded (almost 50% of them in the last five years) and how many employees they account for (40% of them have fewer than 20 employees).

Most companies in the sample (90%) are either *scaling* their customer base (27) or in *commercial* development stages (56). In addition, we have included *developing* DTEs working in new innovative concepts, trading IoT data (e.g., Streamr or IOTA) and AI/ML data (e.g., Skychain), or integrating blockchain in decentralized architectures (e.g., Lemochain, Madana, and Dataeum). Finally, we chose not to include any *open data* providers or repositories. Despite their popularity among researchers, we wanted the survey to focus on DTEs aiming to offer paid data products in the market.

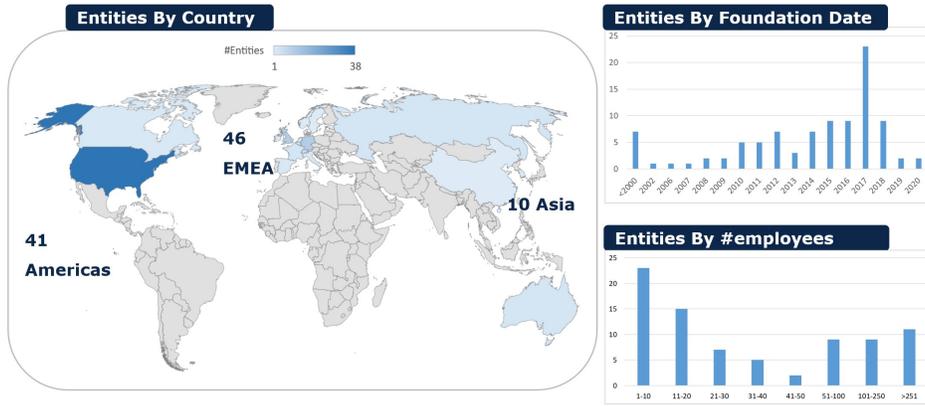

Figure 2: Summary of entities included in the survey

Appendix A thoroughly explains the methodology we followed, including the set of questions we set out to answer, how we gathered and analyzed the information, and some limitations of our study. In addition, appendix B shows the list of the entities included in the scope of this paper.

### 3.2 Classifying DTEs depending on their customers

First, we found that the business models of DTEs heavily depend on who they consider their customers to be, which in turn depends on which side of the chain they approach data trading from. *Data management systems* (DMSs) focus on managing the information an enterprise or individual owns. Conversely, traditional *data providers* (DPs) focus on data consumers, and conceal data owners and often even their partners when selling their products. Whereas the former approached data trading in order to allow secure data exchanges within an organization or to authorize third parties, the latter implemented data trading platforms to complement their existing products or services with those of third parties. In addition, *data marketplaces* (DMs) were conceived from the beginning as two-sided platforms dealing both with buyers and sellers.

Within the scope of the survey, we included 38 DMs and 24 DMSs. As regards DPs, they often provide their products in commercial DMs, and we managed to list 2,015 of them selling their products in a sample of nine public or semi-private DMs. Hence they are by far most frequent business model within DTEs. Since the way they operate is often similar, we took a diversified sample of 35 to understand how they deliver data and how they price their services or products.

We learnt that DTEs do not necessarily implement every layer or process in Fig. 1. For instance, some platforms (e.g., Cybernetica) only implement a secure data exchange lacking any pricing or bidding functions, which are left out so that both involved parties can agree upon them. Unlike DPs, DMs or DMSs, such data trading *enablers* exclusively provide their services to other DTEs.

### 3.3 Data trading business models

Having introduced the first level of business models, in this section we dive deeper into the differences between them, and present some sub-models that group together entities of specific characteristics that belong in the same business model (see Table 1).

In the previous section, we introduced the concept of enablers. We found PIMS and DM enablers that provide a heterogeneous set of open and flexible solutions and platforms, on top of which fully functional platforms can be built. The range of solutions includes, for example, anonymizing personal information (AirCloak), providing an homogeneous anonymized identity to buyers (Datavant), or facilitating secure exchanges (Cybernetica). When it comes to charging and billing, enablers usually charge for transactions (e.g., calls to the API, volume of data processed, etc.). Even though some enablers focus on specific types of data (e.g., IoT-related, AI/ML models, personal data), or industries (e.g., health), we did spot some general-purpose enablers as well (e.g., those providing secure data exchange of distributed data between different entities).

With regards to entities providing full-fledged seller-to-buyer solutions, Tab. 2 summarizes the characteristics and differences of the business models we identified. In the next sub-sections, we provide examples for sub-models of DPs, DMs and DMSs and further explore their differences.

Table 1: Taxonomy of data trading business models

|  | **Data Providers (DP)** | **Data Marketplaces (DM)** | **Data Management Systems (DMS)** |
| --- | --- | --- | --- |
| End-to-end DTEs | Service Providers | General-purpose DM | Embedded DM |
|  | Data Providers | Niche-DM | PIMS |
|  | Private marketplaces (PMP) |  | Survey PIMS |
| Enablers |  | DM enablers (DME) | PIMS-enabler |

*3.3.1 Data Providers.* DPs are entities which provide *data* as a product, be they raw or enriched data, access to information through a graphical user interface (GUI), or information contained in reports to third parties. They usually combine data from different sources (e.g., from the public internet, from partners they collaborate with, or that which is bought from another data provider) to enrich their products and add value to their offer.

*Service Providers* (SPs) are entities providing digital services to end-customers, be they individuals or enterprises, based on data they own, or on that which they collect from the internet, or acquire from third parties. Examples of SPs are Clearview.ai, a company that provides data identification based on pictures of people publicly available on the internet, or Factual, which offer marketing insights based on the movement of people. The boundaries between SP and DP are often blurry: are not personal identifications provided by Clearview.ai or insights by Factual data in the end? To make differentiation even more difficult, some SPs often act as a DP, when they also offer access to their raw data as a product in addition to other services.

From our point of view, supply side platforms (SSPs) and demand side platforms (DSPs) are SPs related to the online marketing industry. SSPs allow publishers and digital media owners to manage and sell their ad spaces, while DSPs allow advertisers to buy such advertising space, often by means of real-time automated auctions. Related to this is the fact that data management platforms (DMPs) refer to audience data management systems that allow advertisers to enrich their audience data with that provided by the DMP. Some marketing-related SPs (Liveramp, Lotame, Openprise, among others) are integrating *private marketplaces* (PMPs) into their DMPs to allow secure exchanges, trading and monetization of audience data from trusted partners (among them the so-called *data brokers*) within the platform. PMPs are frequently an add-on to DMP subscriptions, and therefore can only be accessed by their users.

Despite the fact that the term PMPs often refers to marketing-related SPs, similar business models also flourished in trading geo-located data (Carto, Here), as well as business technographic data (Crunchbase), and financial (Factset, Quandl, Refinitive) data. Such PMPs provide users of these services platforms with relevant second-party and third-party data to enrich their own. As opposed to public or semi-private DMs, data exchange in PMPs is a *private functionality* of DPs or SPs that complements their main value proposition, and hence is only accessible by their customers on the buy side, or authorized data partners on the sell side.

Interestingly, as well as directly commercializing their services through their websites, DPs and SPs also make use of DMs to advertise their services, provide access to free samples of data, or offer specific data products. We found that 45% of data brokers (like Experian, Acxiom or Gravy Analytics) that offer their products marketing-related PMPs (Liveramp, TheTradeDesk or LOTAME) commercialize their products in other DMs such as AWS or DataRade, too. This is also the case with providers such as RepRisk, Equifax or Arabesque S-Ray, who make use of the aforementioned financial-related specialized PMPs.

*3.3.2 Data Marketplaces.* DMs are mediation platforms that put providers in touch with potential buyers, and manage data exchanges between them. Such exchanges usually involve some kind of economic transaction, as well, either through payments in fiat currency or in a cryptocurrency often created and controlled by the platform. DMs are either public - i.e., open to any data seller or buyer - or semi-private, meaning any seller or buyer is subject to the approval of the platform in order to be allowed to trade data. Furthermore, DMs often deal with data categorization, curation and management of metadata to help potential buyers discover relevant data products.

Whereas *general-purpose DMs* like AWS, Advaneo or DataRade trade any type of data, *niche DM* are focused on certain industries (martech, automotive, energy) and on certain types of data (spatio-temporal data, or that coming from IoT sensors). By analyzing the foundation date, we spotted a clear trend towards real-time data streaming marketplaces to harness the potential of IoT (IOTA, Terbine), and DMs specialized in training AI/ML models (Skychain, Ocean Protocol).

Table 2: Summary of business models trading data

|  | Data Providers (DP) | | Data Marketplaces (DM) | | Data Management Systems (DMS) | |
| --- | --- | --- | --- | --- | --- | --- |
| **Concept** | DP/SP | PMP | General-purpose DM | Niche DM | Embedded DM | PIMS |
| **Data exchange** | Public, semi-private, private | Private | Public / Semi-private | | Private | Public / Semi-private |
| **Scope** | Focused | | Diversified | Focused | | |
| **Type of data** | Any | Specific data to be used within their service / platform | Any | Industry or type-specific | Data exchanged within the system | Personal data |
| **Roles / Players interacting** | Partners, Customers | | Sellers, buyers | | Owner, requester | Users, data Providers, buyers |
| **Gets data from** | internet, self-generated, partners, users | Partners, Data providers | Data providers | Data providers, self-enriched | Data providers | Users, Data providers |
| **Provides buyers with** | API, datasets | API, access to data through the system | API, datasets | | API, Access to data through the system | API, Key to decrypt data |
| **Owners get access through** | Partnership | Partnership, the service platform | Web-services | | Data Management platform | Mobile App Web services |
| **Buyers get data through** | Web-services, APIs | Web-service, the service platform | Web-services | Web-services, APIs | Data Management platform | Web-services, APIs, compatible systems |
| **Type of platform** | Centralized | | Centralized or decentralized | | Centralized | Decentralized |
| **Access pricing for buyers** | Subscription, pay for data | Included in the main platform | Predominantly free. Some freemium, subscription and data delivery charges | | Add-on to the data management Platform | Pay for data |
| **Access pricing for sellers** | Partnership (when applicable) | Partnership, time subscription | Predominantly free. Some freemium subscription, and revenue-share charges | | Subscription to the platform | Free |
| **Data pricing schemes** | Fixed one-off, subscription, customized, volume-based | Subscription, specific (e.g., CPC, CPM, …) | Fixed one-off, subscription and customized | Customized, volume/usage-based, fixed one-off | Open | Open, bid by buyer |
| **Control of data pricing** | Platform | Platform, buyers | Platform, providers | | Open | Users, Platform |
| **Payment method** | Fiat currency | | | Fiat, token | Open | Token, fiat currency |

This is also a recent line of research within the scientific community (see Sect. 5).

Data sellers and buyers are often invited to subscribe for free to the platform. However, some platforms charge for freemium subscriptions or charge IaaS-like fees for the delivery of data. A few of them opt for charging sellers according to the money they make through the platform, either through commissions or revenue sharing.

In addition, buyers oftentimes pay the DM for data. Both the data seller and the DM are in charge of setting the prices for data products - in most cases one-off charges for downloading or gaining access to datasets, or periodic subscriptions to data feeds in general-purpose DMs. Conversely, niche DMs more frequently resort to volume or usage-based charging for APIs, and price customization depending on who the data buyer is.

Finally, *data marketplace enablers* (DMEs) are platforms and services that provide partial functionality on top of which full-fledged end to end DMs can be built. As an example, Ocean Protocol provides marketplace enabling functionality for AI/ML data trading. Third parties may use their services to develop end to end DMs providing end to end services and processes as stated in Sect. 2. For example, GeoDB and Decentr are examples of DMs that use Ocean Protocol as a DME.

*3.3.3 Data Management Systems.* Data Management Systems (DMSs) aim to collect, organize, store, combine and enrich information within an organization (enterprise DMS) or, more recently, personal data from individuals (PIMS).

DMs have also flourished within the scope of enterprise DMS. They are offered as an add-on that allows to carry out secure data exchanges within an organization, and to enrich its corporate information base by acquiring data from second or third-party providers. Such *embedded* DMs rarely include full marketplace functionality, but rather restrict themselves to securing data exchanges, and controlling the delivery and access to data assets within the walled-garden of information under the control of each customer. Some of them charge IaaS-like fees for delivering data, and a recurring subscription fee to authorized sellers.

On the other hand, PIMS look to empower individuals to take control of their personal data. They leverage recent data protection laws (e.g., GDPR, CCPA) so as to let users collect personal information controlled by digital service providers, exercise their erasure or modification rights as granted by law, manage permissions of mobile apps to give away personal information, and manage cookie settings, among other things. They eventually aim to establish a single point of control for individuals to manage their PI.

In addition, some PIMS seek their users' consent to share their PI with third parties through the platform and to obtain a reward for doing so. Almost half of the surveyed PIMS include marketplace functionalities, and focus on trading personal data for marketing purposes such as user profiling and ad targeting. Therefore, they leave data subjects (owners of their PI) and data providers to negotiate fees for consenting to get access to data. This way they become personal data brokers, letting users monetize their data, and controlling who has access to it and for what purposes.

Recently, health-related PIMS (Longenesis, HealthWizz, MedicalChain) specialize in managing healthcare-related information of their users. We found that health-related PIMS often resort to DLT (e.g., blockchain) to provide additional security to such sensitive data, and comply with a very strong sectorial regulation.

*PIMS enablers* offer partial PIMS features that may be used by other PIMS, as well. For example, SayMine facilitates users to exert their rights to erase or reclaim their PI held by data providers. Other enablers omit the marketplace functionality, and simply provide users with a secure exchange of PI which they consent to, without providing any economic layer.

Finally, *survey PIMS* aim to facilitate targeted marketing surveys among their users, leveraging information about their profile to offer an accurately targeted audience, and rewarding users for participating in the processes. Citizen.me, ErnieApp or People.io are examples of such a business model.

As opposed to enterprise-DMS, PIMS are more decentralized platforms that leverage the users' devices to store information, and they are always offered for free to individuals. Some charge one-off fees, subscription, or data delivery fees to potential data buyers.

## 4 DEEPER INSIGHTS INTO DATA TRADING

Having characterized the sample of entities and their business models, this section takes a closer look at the results of the survey and their interpretation. In each section we tackle questions related to similar topics, grouped as follows:

(**4.1**) What kind of data is being traded?

(**4.2**) How is data being priced?

(**4.3**) Which payment method and currency are used in data transactions?

(**4.4**) How do platforms charge users for their services?

(**4.5**) How do entities trade data?

(**4.6**) How do entities store data?

(**4.7**) How can the data buyer see or test the data before it is transacted?

(**4.8**) What kind of security measures are taken?

### 4.1 What kind of data is being traded?

As Figure 3 shows, very different kinds of data are being traded in the market. In fact, DTEs are often classified based on the kind of data they trade. For example, we will talk about *marketing DPs* or *marketing PIMS*, meaning DTEs specialized in providing data or managing and trading personal information for marketing-related purposes. We will also discuss the aforementioned *general-purpose* DTEs which don't aim to trade any specific kind of data.

Even though most entities are clearly targeting the business market, no restriction prevents individuals from also acquiring data. DTEs usually target specific industries, and often specific departments within their business customers. As shown in Fig. 3, data may come from different sources, such as PI owned by individuals, data related to companies, industries, measurements from sensors, etc.

Figure 4 shows a breakdown of the kind of data traded by DMSs (in blue), DMs (in orange) and DPs (in grey). There are notable differences in what kind of data entities do trade depending on their business model.

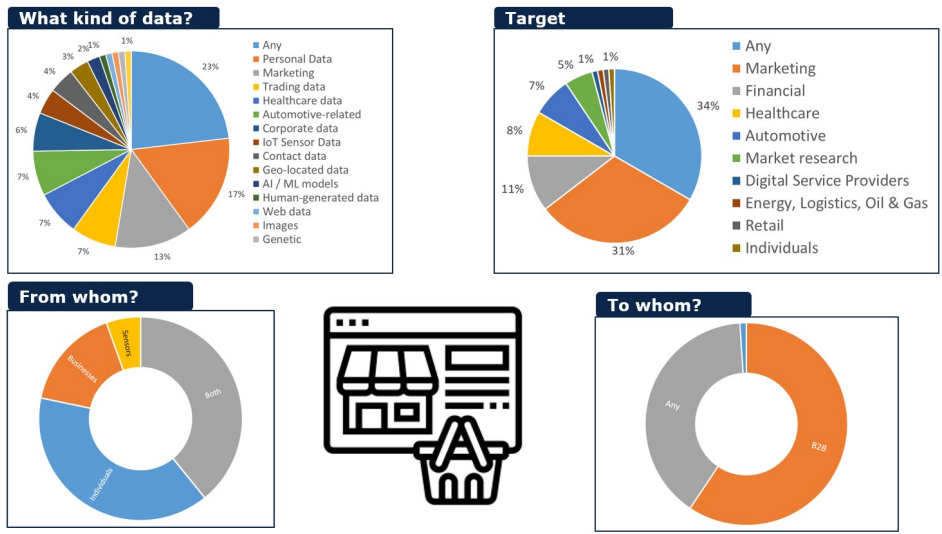

Figure 3: Classification of data being traded by surveyed entities

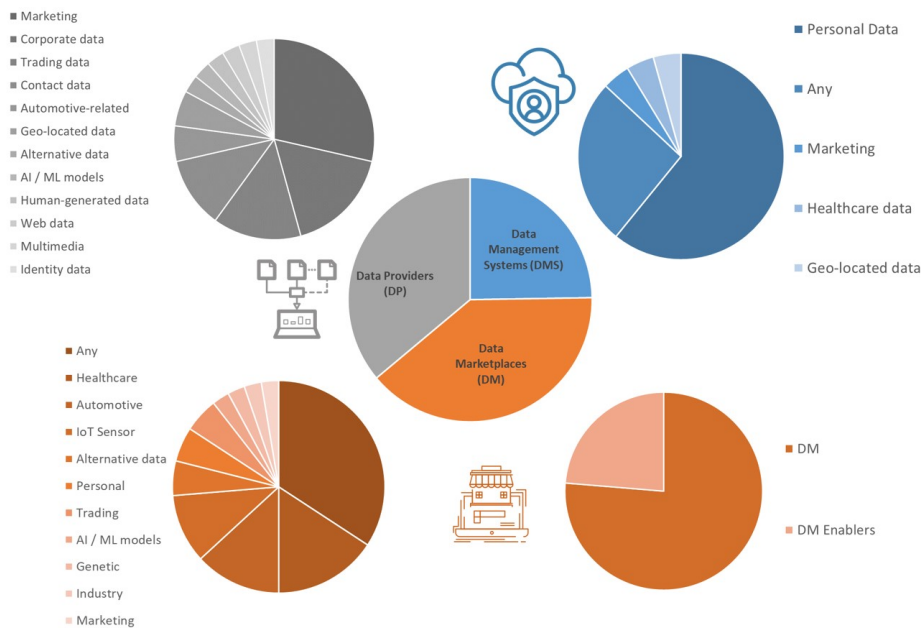

Figure 4: Business models found within the set of surveyed entities, and the kind of data they trade

In general, DPs are specialized in a market *niche*, either a specific type of data or a customer segment. Only one DP (Quexopa) is publicly focusing on collecting and delivering data for a certain region (Latin America). Even though the range of data they deal with is diverse, it turned out that most DPs in our sample are related to marketing data, corporate, contact or financial data.

Within DMS, PIMS focus on personal and healthcare-related data, whereas business-oriented DMS are usually designed to trade different types of corporate data.

With regards to DMs, at least 13 of them are *general-purpose* and trade *any* kind of data, whereas *niche* DMs deal with healthcare, automotive, IoT-related, trading or alternative investment data, as the pie chart at the bottom of Fig. 4 shows.

Focusing on *general-purpose* DMs, we carried out a deeper analysis, drilling down to the level of data products, to better understand what the categories of data most frequently offered in those markets are. For that purpose, we gathered public information about almost two million data products from ten public or semi-private DMs that fulfill the necessary criteria for a measurement study, namely AWS marketplace, DIH, Advaneo, DataRade, Knoema, Snowflake, DAWEX, Carto, Veracity, Crunchbase and Refinitiv.

We collected the categories and tags of each data product and matched the types of products in different DMs to arrive at a common classification hierarchy. Given that each DM uses different categories and tags[1], we matched them at a very high level to avoid as many ambiguities as possible.

As such, Fig. 5 presents the most frequent data categories of data products in general-purpose DMs. The pie chart on the left includes free and paid data products, whereas the one on the right includes only those that are paid (10,860 products). Similar to Fig. 3, '*Marketing*' and '*Economy and Finance*' fall among the most popular categories for paid data products. Moreover, the presence of '*Geography and Demographics*' and '*Geospatial*' data marks the importance of geo-located data in the sample, as well.

Other interesting takeaways from this analysis are that most data products in general-purpose DMs are made available for free, and that some DMs such as DIH, Advaneo, and Google Cloud Marketplace lack any significant offer of paid products. We observed that free data products in commercial DMs are either open data collected from open data repositories, or data samples uploaded by DPs.

Surprising though it may seem in the case of entities whose aim is to make profit, DMs like DIH or Advaneo collect and link open data made available by authorities or public institutions. Metadata for these open datasets is often scarce, hence the large number of them labelled as '*Other*' in the pie chart on the left of Fig. 5. The only business rationale we found for this phenomenon is that such a vast amount of data may serve as a '*hook*' for sellers and buyers, and a complement to third party paid data products.

Moreover, we found that DPs are making use of public DMs to upload outdated samples of their products so that buyers can play around with them and get to know how useful the whole data product would be for their purposes, before eventually triggering its acquisition. This practice would indeed be interesting for DMs, provided it was they who effectively sold the corresponding paid product after the trial. However, some DPs were actually reported to upload only sample products, which refer buyers willing to close any data transactions to the DP's website. In such a situation, the host DM merely acts as a showcase for the DP's data products and consequently loses control of any resulting data transaction, which in turn threatens its business model.

We identified 25 DMs specialized either in certain types of data (e.g., AI/ML, or IoT sensor), or in specific industries (e.g., martech, automotive, or energy) compared to 13 public *general-purpose* DMs. In light of the above and the fact that DPs are, by definition, focused on their area of expertise, we can conclude that specialized DTEs substantially outnumber general-purpose DTEs nowadays. However, it is still too soon to say whether general-purpose or *niche* DMs will succeed. Owing to the heterogeneity of products in the market, there might be cases where one type of DM may be preferable to the other, and vice versa.

## 4.2 How is data being priced?

In general, DTEs charge data buyers for the products they acquire, but how do platforms price data that is exchanged in a transaction? Which stakeholders are involved in setting such prices? We found that 25% of DMs do not provide any clear explicit public information about how data pricing works on their platform. The remainder of DMs and PIMS are somewhat flexible in the pricing scheme and allow their users to choose between different mechanisms. To complement our analysis at entity level, we analyzed how data products are being priced in ten general purpose marketplaces [6].

Figure 6a provides a comprehensive summary of what the most widely adopted pricing mechanisms are, namely the following:

- **Fixed price**. Buyers pay a lump-sum as a one-off for a data product, or a **fixed subscription** charge for accessing a stream or service for a period of time. Most entities providing information about prices (57%) support transactions using fixed prices, and these are by far the most frequent solution adopted by DMs and the most widely used scheme for data products.
- **Volume-based**. Price is fixed depending on the volume of information that is downloaded or accessed. Some DPs do directly provide unit prices for data, often with volume discounts. In particular, this mechanism is popular among contact data providers that charge per contact or lead. Some DMs, like Otonomo, which sells data points to automotive app service providers, also charge according to the volume of downloaded data. Moreover, we learnt that data product prices depend heavily on their volume, even though this is not explicitly stated in the price when analyzing data product prices [6].
- **Usage-based**. DPs providing access to data through APIs often charge depending on the number and type

---

[1]Whereas AWS marketplace defines ten different industry-based categories, DataRade maintains a complex hierarchy of more than 300 categories, let alone tens of other use cases.

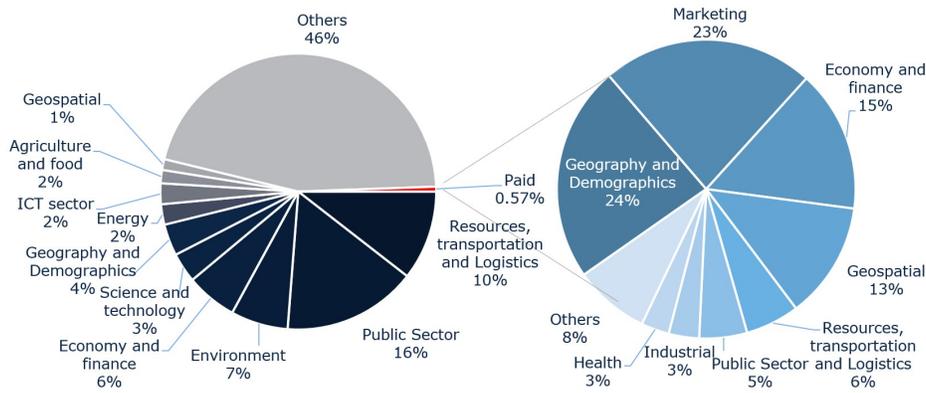

Figure 5: Most popular categories in a sample of general-purpose data marketplaces

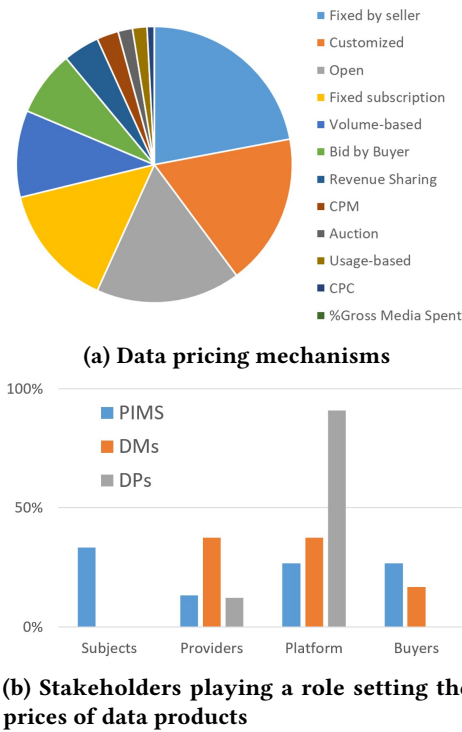

(a) Data pricing mechanisms

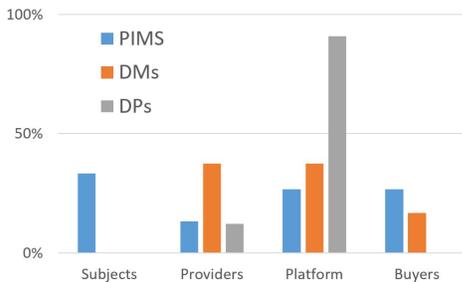

(b) Stakeholders playing a role setting the prices of data products

Figure 6: Data transaction pricing by DTEs in the survey

of calls, often with volume-discounts, as well. Moreover, they usually sell products segmented into different tiers, each of them allowing a maximum number of calls in a certain period of time (usually a month) for a fixed price.

- **Bid by buyer**. Buyers place bids that must be accepted by sellers for the transaction to take place. This way the seller avoids setting an upfront asking price. 12% of DTEs offer this possibility to users, which is more frequently used by PIMS.

- **Customized**. Price is set by the seller case by case, and depends on who the buyer is and what the data is intended to be used for. In general, a transaction starts with the seller asking these questions to the potential buyer before any price is disclosed, which allows the personalization of both data products and their prices, and opens the door to (and triggers concern over) potential price discrimination. DPs were found to be the business model that most often resort to this scheme when pricing their data or services.
- **Free**. Buyers are able to get data for free (i.e., there is no transaction price), sometimes because access to data is included as part of a subscription to the platform (e.g., Carto).
- **Open**. The DTE does not provide any mechanism to set prices, and it is left to buyers and sellers to agree on them. This is the natural approach of enablers, which allow one or more of the aforementioned pricing schemes to be implemented on top of their services (e.g. Ocean Protocol), or simply don't deal with the economic terms of the transaction (Meeco).

In addition, we identified other mechanisms being used in specific contexts.

First, MyDex claims to charge transactions using **revenue sharing**: when a buyer purchases the rights to access the PI of a user, the platform claims its rights to 4% of the revenues that such a buyer makes on the platform from that individual. Revenue sharing requires downstream control of the use of data, which discourages its indiscriminate implementation in charging for data transactions. Digi.me, a PIMS enabler, also mentions this pricing scheme. Although innovative in terms of pricing data, its feasibility is still to be proven: would PIMS be able to control how much money data-buyers are making from each individual user and charge for personal data according to this?

In a different setting, data partnership agreements often resort to revenue sharing when charging sellers for data transactions. Not only is it being widely used by DPs, but also by DMs (e.g., DataRade) or PMPs (e.g., TheTradeDesk). It is important to note that revenue sharing is used to divide data revenue among entities taking part in a transaction rather than to set the price of data in this context.

**Cost per mile impressions (CPM)**, **cost per click (CPC)** and **percent of gross media expenses** are specific to PMPs of online advertising platforms (e.g., LiveRamp, Oracle or Kochava), thanks to their end-to-end control of online ad campaigns.

Finally, **auctions** are very popular price setting mechanisms in other fields, and they are widely used in online advertising where advertisers bid in real time to show their ads to a user browsing a certain webpage [48]. Nonetheless, they are not so common when selling data, due to its non-rivalrous nature. Even though some works of research have already defined a whole family of auctions that artificially creates competition among interested buyers [27, 28], we found only one enabler (Ocean Protocol) that mentions auctions as a potential mechanism to set the prices of data products.

When it comes to the question of who is in charge of setting the price of data products, it is clear that this also depends heavily on the business model as Fig, 6b shows. Whereas DP tightly control the price of their data or services, PIMS give more control to their individual users (the actual data subjects), and usually let them agree with buyers on personal data transaction prices. Although DMs usually play an active role in the process of setting the prices for data products on their platform, they always do it in conjunction with sellers (DPs). In fact, some of them (Dawex, Battlefin) charge for it, and offer specific professional services to buyers as an add-on covering the development of a tailored data-monetization strategy and advisory in setting the prices for their data products.

## 4.3 Which payment method or currency is used in data transactions?

Another relevant and interesting topic regarding data transactions relates to the currency used in data payments. Whereas data providers have traditionally been charged for their services in fiat money (dollars, euros, etc.), 50% of surveyed PIMS and 40% of marketplaces are using cryptocurrencies instead. The benefits they seek by using this alternative include an increased speed of transfers, a higher availability if compared to going through banks or establishments, and a greater liquidity. Real-time data exchanges like the ones trading with IoT sensor data are broadly opting for cryptocurrencies when it comes to liquidate payments.

## 4.4 Do data trading platforms charge users for accessing their services?

DTEs that operate as platforms do not only charge users for the data they consume, but for other concepts such as delivering data, or even just for gaining access to their services. Again, such additional platform charges vary greatly between business models.

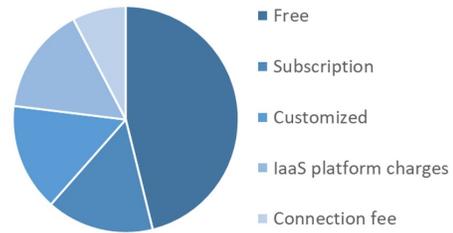

**Figure 7: PIMS charges and pricing to buyers for gaining access to the platform**

In general, PIMS are free for data subjects. On the one hand, this makes sense since they provide the platform with PI to work with, and also make the promise of increased privacy and data protection more appealing. On the other hand, it raises concern over the profitability of users who are unwilling to share their data and are using PIMS for such purposes. Data buyers, who are also usually welcome and free to join the platform, often just pay for the data they acquire. In some cases, potential buyers are asked for a one-off connection fee or charged a periodic subscription (see Figure 7) to get access to the platform. Finally, some platforms demand details about the buyer signing up to the platform to customize such access charges.

Conversely, charging buyers and especially data sellers for access is more usual in the case of DMs (see Figs. 8a and 8b), either through:

- time-based subscriptions, often using a freemium model;
- revenue sharing, where the platform keeps a percentage of the total sales;
- one-off fees to connect to the system.

A few *niche DMs* (Otonomo) and most PMPs offer partnership models to big data sellers, an *ad hoc* agreement to share data frequently used by DPs. Interestingly, a niche DM (Caruso) requires a partnership agreement to be signed by buyers, which requires their participation as shareholders if they are willing to use the platform.

Regarding PIMS and DM-enablers, they welcome full-fledged PIMS to use their technology and usually charge pay-as-you-go IaaS/PaaS-like fees based on the number of API calls or the volume of data they deliver. Some DMs (e.g., AWS or Snowflake) do charge data shipping fees to both parties, too.

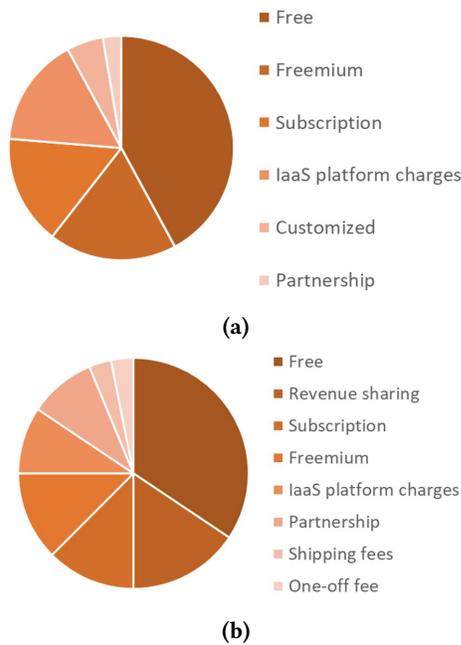

**Figure 8: DM charges and pricing to (a) buyers and (b) sellers using the platform**

## 4.5 How do entities trade data?

Some specific characteristics of data, in particular its zero-cost replicability and its inherently combinatorial value, make this attractive asset considerably more difficult to be priced and safely traded [50, 56]. In economics, a good or service is called *excludable* if it is possible to prevent consumers who have not paid for it from having access to it. In addition, data is non-depletable and hence a *non-rivalrous* good: selling data to customer A does not prevent the owner from selling it to customer B.

Entities trading data aim at somehow making it *excludable* and therefore a club good. Indeed, this is a key challenge in building a flourishing economy around data. This section provides some additional insights about how entities in our survey are attempting to achieve this goal. In the following subsections, we answer questions regarding where entities take data from, what they provide buyers with, and how users - both from the buy and sell sides - gain access to data.

Since the conclusions are very different for DPs, DMs and PIMS, we present them separately in subsections.

*4.5.1 Data Providers.* As Fig. 9 shows, DPs leverage the internet and access to exclusive self-enriched data-sources to provide buyers with access to data either through APIs or bulk downloads, and preferably through web-services or specific applications.

Note that they are not meant to be two-sided platforms, but players oriented to provide their data or their data-driven services to their customers. Should they require proprietary information from third parties, they establish partnerships or bilateral agreements to access such exclusive information, which they eventually enrich and resell. Therefore, DPs control the whole go-to-market process, and conceal the identity of their partners and the sources of their information, unless disclosing them adds any value (e.g., credibility) to their business.

As an exception, PMPs integrated in data-driven services (e.g., spatio-temporal data marketplaces integrated in GIS cloud SPs) allow third-party DPs to sell data within their platform. Unlike DMs, PMPs carefully select authorized DPs, who often sign private partnership agreements with them. Moreover, they deliver data to be used within their system or services, and only to their users, which is why such marketplaces qualify as *private*.

*4.5.2 Data Marketplaces.* Figure 10 shows that DMs collect and sometimes enrich or combine data from different DPs (sellers), who have signed the DM's public terms of use. Similar to DPs, data is often delivered to buyers as a bulk download or through APIs. Although some of them restrict delivery methods to get access to data through their platforms (e.g., AWS marketplace offers access to data stored in Amazon S3 services). they often resort to APIs and web services for buyers and sellers to manage their transactions and data within the system.

*4.5.3 PIMS.* PIMS collect and manage personal data from individuals. Such data is either shared by their users or stored in their devices. PIMS help users retrieve their PI from third parties like social networks or e-mail services, which they call their *data providers*. Then, PIMS may sell such personal data to potential buyers with the owners' consent. Most frequently, PIMS users use a mobile application to get access to the service, which also allows them to manage their consent to share their PI and monitor data transactions. Most of them provide buyers with APIs or web services to gain access to data. As opposed to DPs and DMs, some PIMS ask entities willing to acquire and gain access to data they manage to integrate their apps and systems (MyDex, GeoDB, DataWallet).

PIMS deliver data in technologically innovative ways. For example, some of them provide access to encrypted data streams by sending temporary keys that are revoked once the subscription expires. Some of them resort to hashed temporary URLs to provide buyers with access to data for a certain period.

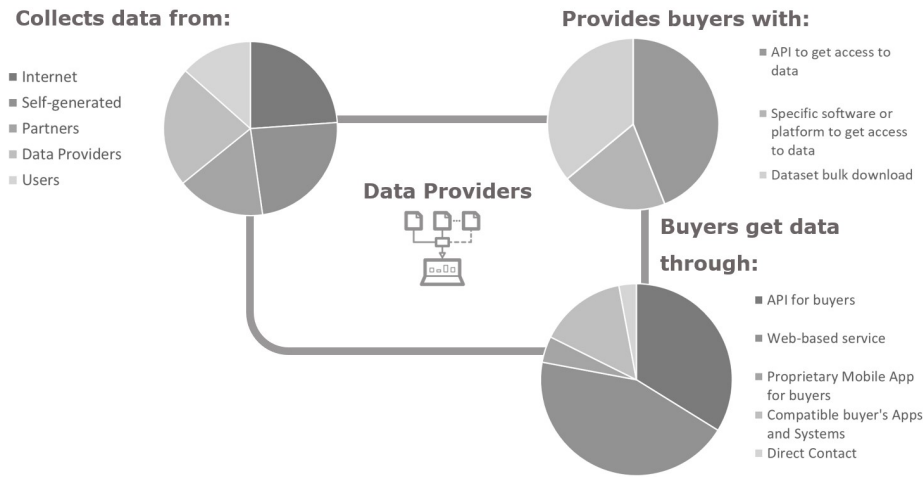

Figure 9: How do data providers work?

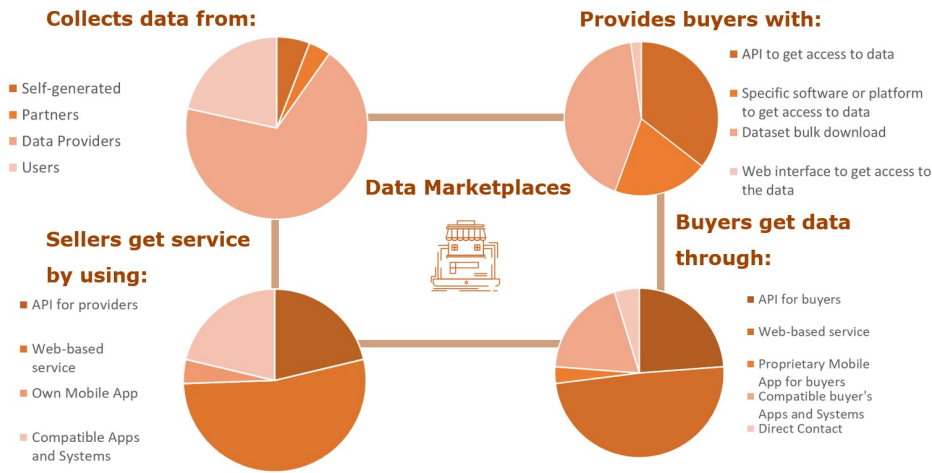

Figure 10: How do surveyed data marketplaces work?

Some PIMS still do not provide an automated platform for buyers to get data and results, but instead they negotiate directly with buyers, and generate the data to be shared with data buyers case by case.

### 4.6 How do entities store data?

PIMS usually opt for a more decentralized architecture by leveraging data subjects or providers to store and process users' data. With some exceptions, they avoid making copies of PI, and usually information is retrieved from the users' personal data storage. On the contrary, DPs and DMs have traditionally preferred a centralized information storage architecture. Some of them (e.g., Advaneo) rely on DPs to store information that is delivered (either sent, or which access is given to) through *secure connectors*, once DPs acknowledge the DM has consented the transaction and secured the payment.

We noticed the existence of a trend towards decentralization of data storage and exchanges. Figure 13 shows how DTEs opted for a centralized or decentralized architecture based on their foundation year. Most recent startups and novel architecture proposals resort to decentralized architectures regarding the storage of both data products and transactional data. In fact, DLTs are increasingly being used to store transactional or management data related to data trading. Due to the high cost of storing data in a DLT, it is not yet being considered as a alternative for storing data for sale, except for specific concept models and developing prototypes related to healthcare (BurstIQ, MedicalChain),

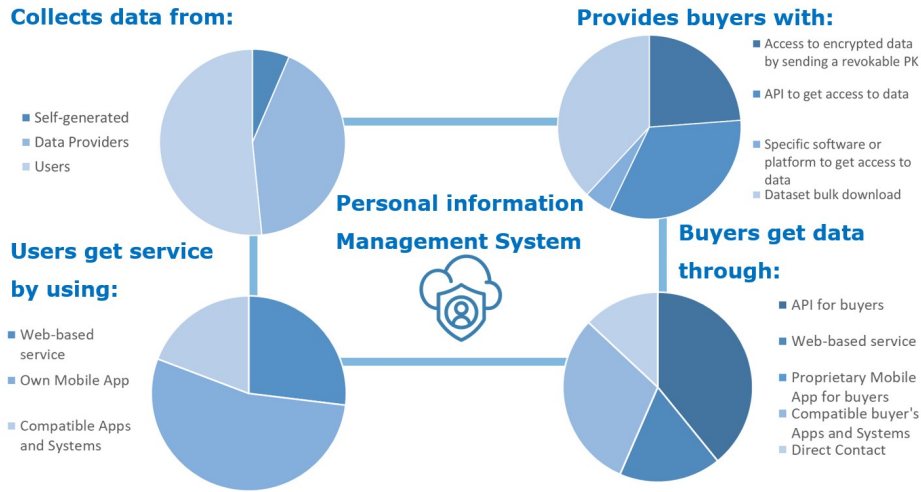

Figure 11: How do PIMS work?

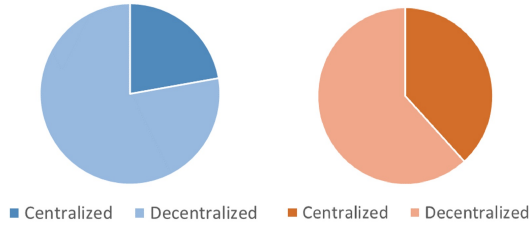

Figure 12: Data management architecture

PI (Dataeum, and Datum, which has already closed) and automotive (AMO), whose feasibility is yet to be proven.

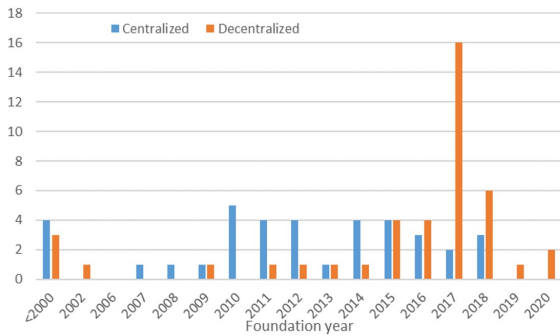

Figure 13: Data management architecture in time

### 4.7 How can the data buyer see, customize or test the data before it is transacted?

Data buyers are not able to realize the value of a piece of data unless they see or manipulate such data. This fact, which is often known as Arrow's information (or disclosure) paradox, often deters data trading. Consequently, a big challenge for DTEs is finding ways to reduce the uncertainty of buyers. We addressed this issue by attempting to understand whether they provide mechanisms for buyers to test, see or customize the data before purchasing it, and whether this reduces the risk of uncertainty.

As a result, we found that 67% of entities provided an answer to this question on their websites, which reflects this is indeed an important issue for them. They claim to be using one or more of the following mechanisms:

- Publishing or sending in advance **free samples** of data to potential buyers or allowing **free access** to part of the data (e.g., some fields of a structured data base).
- Offering a **trial period** in which to have access to a data feed or subscription-based service.
- Providing buyers with a **sandbox** (Battlefin, Otonomo), a controlled environment that lets them play with real data before bidding for it or making a purchase decision, while ensuring that data is not downloaded nor copied.
- Offering a live **demo** of their services and the data they offer.
- Hosting a **reputation mechanism** by which buyers are able to rank both information and/or data providers.

Free samples are by far the most widely used method to let buyers know a data product in advance. Not only do DTEs offer them on their websites; we also found that a number of DPs publish such samples in public DMs which buyers can download for free.

We found that being able to know the value of data beforehand is not equally critical in all circumstances. On the one hand, it seems to be a key challenge for general-purpose DMs or those trading AI/ML data, due to the heterogeneity of data they trade and the difficulty of anticipating the usefulness of pieces of data before testing them [9, 24]. On the other hand, DPs offering real-time products such as data feeds or streams often allow buyers to cancel subscriptions at any time, so that the risk of finding data useless after subscribing to a feed is therefore limited. In other contexts, PMPs (Lotame, Liveramp), marketing (Wibson, Vetri) and health-related (HealthWizz) PIMS sell data customization services (e.g. customer segmentation) that allow buyers to tailor the characteristics of the data (audience) they are purchasing, thereby reducing the risk of ending up acquiring useless data.

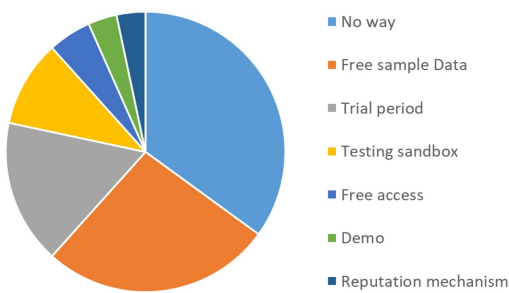

**Figure 14: Mechanisms to provide trust and visibility of data products to buyers**

## 4.8 What kind of security measures are taken?

PIMS and DMs often publish high-level information related to security of data and exchanges in order to gain the trust of potential users, be they buyers or sellers. In particular, PIMS express the greatest concern about users' privacy: most of them include a section dedicated to data security, and answer FAQs to address users' and buyers' concerns in this respect. On the contrary, DPs are generally reluctant to give away any information about security, which is considered an internal policy.

Some of the measures taken by DTEs to secure data include:

**User authentication and identification** and **SSL encryption**, which are widely used. The first is useful to control who is accessing the platform, in which role, and with which permissions, while the second one ensures the security of information while "on-the-move".

**Anonymization or de-identification** of personal information (e.g., through hashing, by Fysical) is used to make identity matching more difficult. This is actively used by PIMS. For instance, Airbloc defines its own zero-knowledge proof identity matching to avoid exposing PI, and by default shares only non-personal information from users.

**Revokable DLT decryption keys**, (DAWEX) or public-key cryptography (Streamr, Datum, Mydex), which are implemented by some marketplaces and allow buyers to decrypt an encrypted data stream or dataset while their contract with the seller is still valid. Such DMs often include additional security measures for storing such keys (Digi.me). Other PIMS rely on a **temporary URL** to provide access to data (Datapace).

**Secure data connectors** (Advaneo, DIH, Databroker), which are used by data exchange platforms to make sure that both parties trade data within the scope of a valid contract.

To protect buyers, some DTEs provide **tamper-proof data through data signatures and message chaining**, which sometimes make use of a blockchain to ensure immutability (Datapace).

Some IoT marketplaces offer sellers a **specific service and software that certifies the origin of data**, which is sometimes bundled with the secure connector and allows some data management functionality (Veracity).

Some DTEs sell data to be used within a walled-garden of their systems and services, and heavily restrict outgoing data flows. For example, *embedded* DMs in enterprise DMS make use of governance and data management functionality to control and manage the access to data stored in the DMP, while also avoiding multiple copies of data. Similarly, PMPs sell data intended to be consumed within their platforms.

Notwithstanding the aforementioned measures, DMs still fail to provide a fully effective solution to avoid data replication. Moving from providing data to providing services has traditionally been the most commonly accepted recipe to mitigate this risk [56]. Therefore, AI/ML niche DMs look to sell model training services [18], rather than bulk data for users to train their models as general-purpose DMs do.

Extending the scope of controlled environments like embedded-DMs might be a means to impose severe barriers to data replication and enhance the control of the access to data. Still it needs to be proven that such a "walled-garden" concept can be scaled and bootstrapped to the entire internet while respecting internet and web governance principles such as openness, standardization, and layering [34].

## 5 DATA MARKETPLACES IN THE RESEARCH COMMUNITY

Recent vision papers state the different research challenges envisaged by the research community [52] when building data marketplaces. They propose high-level architectures, and point to some key research challenges in DM design, namely: data cataloging and discoverability, data valuation and pricing, uncertainty of buyers regarding the outcome of purchasing processes, and revenue allocation and sharing. Moreover, most of the existing solutions naturally assume the winner-takes-all economics of the internet [33], and therefore they aim to become the Google of data in their specific niche of the market. More radical visions introduce the old concept of Information Centric Network and the need to create an overlay standardized infrastructure to securely and fairly handle data across the internet [34].

An important ongoing research effort is focused on marketplaces intended to train AI / ML models. The trend is towards selling trained models instead of data, and price based on the value such processing brings to the buyers [18]. Different value-based data marketplaces have been designed based on this concept, from sellers selecting a price-value from a mix offered by the DM [14], to buyers bidding for data and returning a proportional value in return [1], or collaborative DMs [45].

The pricing of data has also long attracted the attention of the scientific community from different fields [38]. Some authors point to such multidisciplinarity as a main challenge of current research in the area [50]. As a result, different schools are applying disparate tools to set arbitrage free revenue-maximizing prices to data products, often in specific contexts. Such tools include auction design [27, 28], differential privacy [25, 35], pricing of different queries to a single database [13, 31] and quality-based pricing [30, 65, 66]. Other authors focus on personal data within the online ad ecosystem [11, 37, 48], spatio-temporal [4, 5], or IoT sensors data [39].

# 6 OPEN CHALLENGES

Despite its remarkable potential and observed initial growth[2], the market for business-to-business (B2B) data to be used by ML algorithms to improve decision-making is still at its nascent phase. Like all nascent economies, from the oil boom of the 19th century, to the dot.com bubble of the 1990s, and the cryptocurrency fever of the last decade, the data economy faces a yet uncertain future. Regardless of which companies and business models finally succeed, we identified some key, intertwined, open challenges related to increasing the *trust* of data transactions:

(1) First, dealing with *ownership* and fighting against piracy and theft of data is of uttermost importance to ensure trustworthiness in DTEs. This task is even more arduous when malicious players are able to copy and transmit data at zero cost, and the market lacks a sound notion of authorship. Apart from safeguarding our own data from unauthorized access, many other challenges arise when trading involves different entities: how can data trading succeed while someone can obtain a dataset from a DM, process it and then resell it (potentially at a lower price) at either the same or a different DM?

(2) Regarding data economics, a healthy data market requires acknowledgeable neutral references to avoid ending up in a radical and sustained price fluctuation of data products. Assuming ownership is preserved, such market references must deal with the particular characteristics of data as a tradable 'good', namely its non-rivalrousness, its ever-decreasing close-to-zero copy, process, and transmission costs, and its customer-dependent asymmetric value.

(3) Due to the fact that 'data' is an experience good, it is far from obvious for potential buyers to anticipate the value of data samples in certain settings such as AI/ML tasks [9, 24]. Hence, allowing buyers to select data samples that fit their purposes and improve their models or support their decision-making is important in those situations, too. [8].

(4) Related to *data provenance*, computing *fair compensations* for DPs at scale is an additional challenge for DMs. Such compensations must be in accordance with the value they bring to a specific task or buyer, and DMs must account for them and provide enough transparency about the process.

(5) The *current fragmentation of data markets* makes it very expensive for DPs to establish a meaningful presence in every existing DM, and for data buyers to quickly find and compare data and prices across DMs. Regardless, a foreseeable consolidation could take place in the upcoming years and either a new single monopoly or 'niche' data trading champions may arise. Decoupling data and internet services in the value chain is a task so big and difficult that we should not expect that a single company, or a "walled-garden" ecosystem, will solve it for everyone. Instead, solutions must be sought that respect transparent internet and web governance and expansion principles, including openness, standardization, and layering [34].

Overcoming the grand technical challenges posed by radical new technologies will probably require investing in and developing other radical new technology that implements an *effective data provenance* to establish ownership and track the spread of data traded in the market, and *usage-based economics* on top of the aforementioned provenance layer. Fortunately, some existing cutting-edge technologies will decisively help in undertaking such a titanic task, specifically:

- Usable provenance may be built upon new advances on digital watermarking [2, 15, 21, 32, 36], hashing [26,

---

[2]Looking for example at the Amazon Web Services (AWS) Data Marketplace and DataRade.ai, we observed a growth of over 50% in 2021 [6]

- 67], trusted execution [54], and network tomography [12, 58].
- Usage-based economics may be built upon crowdsourcing, cryptography, and blockchain-based Non Fungible Tokens (NFTs) [22].
- Information Centric Network (ICN) principles [3] at its data layer provide an interesting base to handle personal data naming, routing, and in-network storage and replication.

Finally, significant regulatory challenges related to data trading lie ahead, both for competition authorities and *ex ante* regulatory bodies.

Due to their market power, tech companies are increasingly under the scrutiny of regulators both in the US and the EU. In the same way that incumbent telecommunication operators are forced to unbundle assets in the access network and grant wholesale access to new entrants, policymakers are currently evaluating the imposition of some degree of data sharing to dominant tech firms in their effort to balance its market power [10]. However, designing such a policy is more complex in the case of data assets due to its potential harm to privacy and security.

Within the realm of *personal data*, protecting privacy was the main purpose of recent legislation in the EU (GDPR) [59] and the US (CCPA) [44]. New legislative proposals in the EU [61, 62] aim to foster data sharing and '*offer an alternative to data handling practice of major tech platforms*' [60]. Assuming regulatory bodies are able to enforce such regulations, some authors have proposed that individuals are compensated for their personal data when shared with third parties [33], while others suggest that DTEs collecting personal data must be required to act as a fiduciary [20], or even that the mass collection and sharing of sensitive personal data must be banned and prosecuted [63].

## 7 SUMMARY

We have catalogued ten different business models in this paper, based on a comprehensive survey that analyzed 180 entities trading data on the internet. Through this extensive study, it has become clear to us that most of the challenges these entities face have to do with *trust*. On the one hand, sellers express an ambition for absolute control of their data, and demand a strong commitment from DMs that data is not replicated and resold, nor used without their authorization. On the other hand, potential buyers need to test data and know its value before closing a transaction, and certify that information comes from trusted data sources.

Not surprisingly, the most successful market players nowadays are horizontally integrated service providers that protect (rather than share) their most valuable data assets. They exploit data collected either from the internet, their user base, or that which they acquire from partners. This is then combined, processed and used to feed more elaborate services to the end customer. As a result, their business model becomes more and more difficult to replicate.

Owing to the heterogeneity of data and its potential use cases, most entities trading data tend to focus on certain industries and/or types of data, and these days most agreements are private and bilateral. Traditional DPs are being challenged by DM platforms that work both with data sellers and buyers to facilitate data transactions. It is unclear whether there is a one-fits-all solution, and DMs must still prove that their business model is feasible in the long term and how they will monetize the comprehensive directory of data products they provide, while avoiding the threat that DPs sell their data on their own.

More recent niche DMs coexist with general-purpose DMs in the market nowadays, although it is not clear which business model is more convenient when it comes to trading data. On the one hand, niche DMs have clear advantages over general-purpose DMs. First, because focusing on certain data space and leveraging their specific expertise let them provide value-added services both to buyers and sellers along with data sharing. Second, because their platform is adapted to the kind of data they trade, and they concentrate their commercial efforts on attracting a specific buyer segment. On the other hand, niche DMs target a much smaller market, and the concept of a one-stop-shop for any kind of data is arguably attractive.

Unlike public DMs, embedded DMs and PMPs consider data trading more as a functionality add-on to the services they already provide. This has two important competitive advantages. First, they leverage an existing potential customer base on the buy side, which lets them concentrate on finding the right data partners to attract their captive demand. Second, they sell data to be used within a specific environment, which significantly reduces the risk of replication and lets them provide more focused, processed, and thereby more valuable data,

Fighting against the data-for-services dynamics of the internet is the main challenge of PIMS, provided the rights of new data protection legislation are enforced by competent authorities. They are focusing on gaining the *trust* of users to build a minimum viable base, yet their feasibility is still to be proven. Consequently, they are struggling to make themselves known, leveraging an increasing concern around privacy on the internet. Conversely, the variety of existing isolated platforms may undermine the trust of users. A future consolidation may facilitate their task of acquiring users, though it may well turn the odds against them unless they differentiate themselves from the big '*datalords*'- why trust your PI to PIMS instead of internet service providers?

Adopting data *trust* models might be a way to overcome this challenge [20].

Existing DMs decouple data from AI/ML algorithms and models: they do not perform any processing on behalf of data buyers unless such a service is contracted separately as an outsourcing or professional service. Some new proposals aim to sell trained models, and thus accuracy, rather than data for AI/ML tasks. Other emerging DMs are focusing on IoT, and intend to decouple data generation from data management and processing in the stack. In this field, the trend is towards real-time data streaming marketplaces to harness the potential of IoT, and increase the automation and competitiveness of the industry. Finally, new DMs opt for more distributed architectures, for DLT to store and track at least their transactional information, and for cryptocurrencies to speed up payments.

In conclusion, the data economy is a thriving though controversial ecosystem still under development. A huge corporate, entrepreneurial and research effort aims to *de-silo* data and enable a healthy trading of such an important asset, which is key to fully unleash the power of the knowledge economy. In this study we have revealed significant differences between what is working in the market right now and what the market is developing. Through commodifying data trading, the market is moving away from horizontally integrated monolithic siloed data providers, and towards distributed specialized exchange platforms.

## 8 ACKNOWLEDGEMENTS

This project has been funded by the European Union's Horizon 2020 Research and Innovation program under the PIMCITY project (Project No. 871370).

# A METHODOLOGY

The methodology used to carry out the survey consists of the following steps:

(1) **Identification of target companies** trading or making business by delivering data. Companies were identified by either searching the web with relevant key words, or by browsing through relevant articles and papers available on the internet.
(2) **Making a quick first assessment** and classifying companies according to the following basic parameters: type of data they are trading, target industry, type of clients, and business model.
(3) **Formulation of a comprehensive set benchmark questions** covering all the aspects we want to answer in this study, and defining a preliminary set of possible answers to each of them. This was further refined during the research process to generate a taxonomy for presenting the results of the benchmark.
(4) **Carrying out a desktop research** to dive deeper into each specific company, answering to the survey questions in a data sheet, and generating a detailed information dossier about the company for consultation purposes in a latter stage as needed.
(5) **Building the data taxonomy by homogenizing the answers to the benchmark questions** for each company and refining the existing taxonomy of answers that allows the comparison of companies.
(6) **Analysis of the results** of this study, both from a technical and a business perspective, identification of key business models and entities operating according to them.

Several iterations were needed in order to come up with a comprehensive set of data trading entities, and fully understand the current market situation.

## A.1 Questions

Table 3 summarizes the questions considered in the survey, the different answers we found when studying the different entities, and the section of the paper where the results are presented.

In addition to answering the former questions, we gathered some general data to classify each entity, understand its maturity, and measure its popularity. These KPIs include the foundation year, country of origin, companies backing the project, the number of employees, how much money they raised, its AlexaRank and its trend in the last months.

## A.2 Data collection approach and limitations

Data acquisition was the result of a desktop research based on secondary information available on the internet. As a consequence, the survey relies on information that the target entities are directly publishing on their websites, as well as any related material, such as whitepapers, public videos, product brochures and presentations.

In the next section, we present the results of the survey for each of the questions in table 3. Whenever an answer was not found for any question in the case of a specific entity, "N/A" (meaning *not available*) labels were used. In general, this situation is due to either a lack of information when analyzing such entities, or due to insufficient detail of such information to answer the question. We report the percentage of entities for which we have information in each subsection.

Table 3: Survey questions and taxonomy of the results produced in the survey

| Field | Question | Values | Sect. |
|---|---|---|---|
| Type of data | Which kind of data is the entity trading with? | IoT Sensor Data; Personal Data; Geo-located data; Contact data; Marketing; Corporate data; AI / ML models; Human-generated data; Multimedia; Industry; Trading Data; Web data; Automotive-related; Identity data; Healthcare data; Genetic data; Any | 4.1 |
| Whose data? | Who are the data subjects? | Individuals; Businesses; Sensors; Any source | |
| To whom is it sold (B2C, B2B, Any)? | who is provided with data after each transaction? Who is the data consumer? | B2C; B2B; Any | |
| Targets | Who is the target of the entity? In case of B2B business models, which department or specific industry is the company targeting? | Digital Service Providers; Marketing; Market research; Financial; Automotive; Individuals; Energy, Logistics, Oil & Gas; Healthcare; Retailers; Any | |
| Actor(s) setting prices of datasets | Who sets the price of traded datasets? | Providers; Platform; Subjects; Buyers; Open | 4.2 |
| Pricing mechanisms | Pricing mechanisms available for data being sold by the marketplace | Fixed subscription; Bid by Buyer; Fixed by seller; Auction; Customized; Free; Revenue Sharing; CPM; CPC; %Gross Media spent; Volume-based; Open; N/A | |
| Payment redistribution mechanisms | How does the platform redistributes payments to data subjects / sellers? | One-to-one; Contribution-reputation-based; N/A | |
| Data transaction payment | Which payment method and/or currency is used in such transaction? | Fiat currency; Token; Internal credits; N/A | 4.3 |
| Platform pricing policy towards data subjects | How are data subjects charged for accessing the platform? | Free; Connection fee; Time subscription; IaaS platform charges; Shipping fees; Freemium; Open; N/A | 4.4 |
| Platform pricing policy towards data buyers | How are data buyers charged for accessing the platform? | Free; Connection fee; Subscription; Revenue sharing; IaaS platform charges; Shipping fees; Customized; N/A | |
| Platform pricing policy towards data sellers | How are data providers / sellers charged for accessing the platform? | Free; Connection fee; Time subscription; Revenue sharing; Freemium; IaaS platform charges; Shipping fees; Partnership; One-off fee; Sales commission; N/A | |
| Access for providers | How do data providers get access to the platform? | API for data providers; Web-services; Mobile App; compatible DPs' systems; N/A | 4.5 |
| Access for buyers | How do data buyers get access to the platform? | API for data buyers, Web-services, Proprietary mobile app, compatible data buyers' systems, direct contact, N/A | |
| Data sources | Where is data coming from? | internet; Self-generated; Sellers; Data Providers; Users; IoT devices | |
| Data delivery | How does the DM deliver data? | Access to encrypted data by sending a revokable PK, through an API, through a specific software or platform, by training models with the data; dataset bulk download; Web services | |
| Data storage | Where is traded data stored? | Centralized public cloud backend, Decentralized private clouds, Centralized private cloud, Data subject's device, Distributed depending on data provider, Centralized backend, DLT, Centralized backend or Public cloud, Decentralized public cloud, Data subject's device and Centralized servers, N/A | 4.6 |
| Transaction / Management data storage | Where is the information about transaction stored? | DLT, Public cloud backend, DLT or centralized management, Centralized backend, Distributed depending on data provider, N/A | |
| Structure of data | Who determines the structure of data to be stored? Can the user share whatever data they want? | Data owner, Application, Data sellers, Data providers and the platform, Platform, Data providers, N/A | |
| Data preview | How can the data buyer see or test the data before it is transacted? | No way, Free sample Data, Free access, Demo, Trial period, Reputation mechanism, Testing sandbox | 4.7 |
| Data Security Measures | How do PIMS/Marketplaces prevent unauthorized access to data while stored? And while it is being moved? | Encryption and SSL, DLT decryption key distribution, Distribution of PK, Special additional measures for PI, Secure storage for PK, User authentication, DLT data replication and immutability protection, Distributed secure data storage | 4.8 |

# B LIST OF ENTITIES INCLUDED IN THE SURVEY

Table 4 summarizes the list of entities trading data that were analyzed in depth in the survey, including their business model out of the ones defined in Tab. 1. For bigger companies such as SAP or Oracle, the business model reflects the role of their data trading solutions.

In addition to the former companies, the survey included the following entities: AAAChain, Acxiom, Adcolony, Adelphic, Adform, Adition, AdMaxim, Adobe Advertising Cloud, Adot, AdSquare, adsWizz, adXperience, Algorithmia, Amaxon Mechanical Turk, Amobee, Apervita, Automat, Axonix, Bidtheatre, BigChain, BigToken, Bottos, Bluetalon, CentroBasis, Clearview.ai, Cogito, Complementics, CoverUS, CXSense, Datacoup, Dataguru, DataHub, Datax.io, DataXpand, dbc, Demyst, Evotegra, Experian, Eyeota, Fyber, Hu-manity, Ifeelgoods, IBM, , iExec, Imbrex, ImproveDigital, Informatica Data Exchange, InMobi, LiveIntent, LUCA, Magnite, Mediarithmics, Microbilt, MyHealthMyData, Nielsen, OpenPDS, Opiria Blockchain, Optum Data Exchange, Orderly, OwnData, OwnYourInfo, PickcioChain, PlaceIQ, Pubmatic, Qlik Datamarket, Relevant Audience, Reply.io, Reveal Mobile, ROKU (Oneview), Rubicon project, RythmOne, Smaato, Smartclip, StreetCred, Synchronicity, Tabmo - HAWK, Taboola, TapTap, The DX network, Tremorvideo, Trufactor, Wove, Xandr, XDayta.

After a first quick assessment, we discarded such entities for their subsequent in-depth study and documentation. In particular, we rejected online advertising platforms not offering a private marketplace, concept projects either lacking information or discontinued in time, entities no longer providing service, nor providing any data exchange or data-driven service as such.

Finally, we filtered out some entities those whose business model was already well represented by entities in Tab. 4. For example, we found 2,015 data providers with similar business models, but we only included 35 of them in the survey. We prioritized data or service providers providing clear pricing information. As an exception, we did include every active PIMS we found in the market in order to provide the reader with a thorough overview of this brand-new business model.

Table 4: List of entities included in the survey (accessed: Dec'21)

| Entity | URL | Business model | Entity | URL | Business model |
|---|---|---|---|---|---|
| 1DMC | https://1dmc.io/ | DM | HealthWizz | https://www.healthwizz.com/ | PIMS |
| Advaneo | https://www.advaneo-datamarketplace.de/en/ | DM | HERE | https://developer.here.com/products/platform | PMP |
| Airbloc | https://airbloc.org/ | PIMS+DME | HxGn Content | https://hxgncontent.com/ | DP |
| Aircloak | https://aircloak.com/ | DME | Intrinio | https://intrinio.com/ | DP |
| AMO | https://www.amo.foundation/ | DM | IOTA | https://www.iota.org/ | DM+DME |
| Atoka | https://atoka.io/ | DP | Knoema | https://knoema.com/ | DM |
| AWS | https://aws.amazon.com/marketplace/ | DM | Kochava | https://www.kochava.com/ | PMP |
| Azure | https://azure.microsoft.com/en-us/services/open-datasets/ | DM | LemoChain | https://www.lemochain.com/ | DME |
| BattleFin | https://www.battlefin.com/ | DM | LiveRamp | https://liveramp.com/our-platform/data-marketplace/ | PMP |
| Benzinga | https://www.benzinga.com/apis/ | DP | LonGenesis | https://longenesis.com/ | DM |
| Bloomberg EAP | https://www.bloomberg.com/professional/product/enterprise-access-point/ | DP | Lotame | https://www.lotame.com/ | PMP |
| BookYourData | https://www.bookyourdata.com/ | DP | Madana | https://www.madana.io/ | DM |
| BronId | https://www.bronid.com/ | SP | Meeco | https://www.meeco.me/ | PIMS+DME |
| BurstIQ | https://www.burstiq.com/ | DM | MedicalChain | https://medicalchain.com/en/ | PIMS+DME |
| Carto | https://carto.com/ | PMP | Mobility DM | https://www.mdm-portal.de/ | DM |
| Caruso | https://www.caruso-dataplace.com/ | DM | Multimedia Lists | https://multimedialists.com/ | DP |
| CitizenMe | https://www.citizenme.com/ | Surv. PIMS | Mydex | https://mydex.org/ | PIMS+DM |
| Cognite | https://www.cognite.com/ | Emb. DM | Ocean | https://oceanprotocol.com/ | DME |
| Convex | https://convexglobal.io/ | DM | OpenCorporates | https://opencorporates.com/ | DP |
| Crunchbase | https://www.crunchbase.com/ | PMP | Openprise | https://www.openprisetech.com/ | PMP |
| Cybernetica | https://cyber.ee/ | DME | Oracle DMP | https://www.oracle.com/data-cloud/products/data-management-platform/ | Emb. DM |
| datablockchain.io | https://www.datablockchain.io/ | DME | OSA Decentralized | https://osadc.io/en/ | SP |
| Databroker | https://databroker.global/ | DM | Otonomo | https://otonomo.io/platform/ | DM |
| Dataeum | https://www.dataeum.io/ | PIMS+DM | People.io | http://people.io/ | Surv. PIMS |
| Data Intelligence Hub | https://dih.telekom.net/en/ | DM | Quandl | https://www.quandl.com/ | DM |
| Data Republic | https://www.datarepublic.com/ | Emb. DM | Qiy | https://www.qiyfoundation.org/ | DME |
| DataPace | https://www.datapace.io/ | DM | Quexopa | https://quexopa.io/ | DP |
| Datarade | https://datarade.ai/ | DM | Refinitiv | https://www.refinitiv.com/ | PMP |
| DataScouts | https://datascouts.eu/ | DP | Salesforce | https://www.salesforce.com/products/marketing-cloud/data-sharing/ | DM |
| Datasift | https://datasift.com/ | SP | SAP data marketplace | https://blogs.sap.com/2021/12/13/sap-data-warehouse-cloud-data-marketplace-an-overview/ | Emb. DM |
| Datavant | https://datavant.com/ | DME | SayMine | https://saymine.com/ | PIMS |
| DataWallet | https://datawallet.com/ | PIMS+DM | Skychain | https://skychain.global/ | DM |
| Datum | https://datum.org/ | PIMS+DM | Snowflake | https://www.snowflake.com/ | Emb. DM |
| Dawex | https://www.dawex.com/en/ | DM | Streamr | https://streamr.network/ | DM |
| Decentr | https://decentr.net/ | PIMS+DM | TelephoneLists | https://telephonelists.biz/ | DP |
| DefinedCrowd | https://www.definedcrowd.com/ | DP | Terbine | https://terbine.com/ | DM |
| dHealth | https://dhealth.network/ | DME | The Adex | https://theadex.com/# | PMP |
| Digi.me | https://digi.me/ | PIMS+DME | TheTradeDesk | https://www.thetradedesk.com/us | PMP |
| Enigma | https://www.enigma.co/marketplace/ | DP | USA Sales Lead | sales-lead.org | DP |
| ErnieApp | https://ernieapp.com/ | Surv. PIMS | v10 data | http://www.v10data.com/ | DP |
| Factset | https://www.factset.com/marketplace | PMP | Veracity | https://www.veracity.com/ | DM |
| Factual | https://www.factual.com/ | SP | Vetri | https://vetri.global/ | PIMS+DM |
| Fysical | https://fysical.com/ | DP | Vinchain | https://vinchain.io/es | SP |
| GeoDB | https://geodb.com/en/ | PIMS+DM | Webhose.io | https://webhose.io/ | DP |
| Google Cloud | https://cloud.google.com/marketplace | DM | Weople | https://weople.space/en/ | PIMS+DM |
| GXChain | https://en.gxchain.org/ | DME | Wibson | https://wibson.org/ | PIMS+DM |
| Handshakes | https://www.handshakes.com.sg/data.html | DP | Xignite | https://www.xignite.com/ | DP |
| HAT | https://www.hubofallthings.com/ | PIMS+DME | Zenome | https://zenome.io/ | DM |
| Health Verity | https://healthverity.com/ | DM | | | |